# THEORETICAL STUDY OF TERAHERTZ ABSORPTION SPECTRA AND NEUTRONINELASTIC SCATTERING IN FRUSTRATED MAGNET $Tb_2Ti_2O_7$

V.V. Klekovkina*, B.Z. Malkin

Kazan Federal University, 420008 Kazan, Kremlevskaya str., 18

*vera.klekovkina@gmail.com

**Abstract**

Within the framework of the single-particle approximation, the envelopes of the spectral lines of terahertz absorption and inelastic scattering of neutrons corresponding to magnetic dipole transitions between the sublevels of $Tb^{3+}$ ions in the $Tb_2Ti_2O_7$ crystal, split by the field of random deformations induced by point defects of the crystal lattice upon violation of the stoichiometric composition of the crystal, were calculated.

**Introduction**

Thespectral, thermodynamic and magnetic properties of geometrically frustrated terbium titanate crystal $Tb_2Ti_2O_7$ with pyrochlore structure have been actively studied for over 20 years. Scientificinterest in this crystal is due to the following. Neighboring $Tb^{3+}$ions in $Tb_2Ti_2O_7$are linked by competing ferromagnetic exchange and antiferromagnetic dipole interactions. The Curie–Weiss temperature is −13 K [1]. Accordingto theoretical studies, the antiferromagnetic noncollinear phase of $Tb_2Ti_2O_7$should be observed below temperatures of ~ 1.8 K [2]. However, magnetic ordering is not observed experimentally below temperatures of 0.015 K [3]. Theabsence of magnetic ordering and the formation of a spin liquid state (cooperative paramagnet) in $Tb_2Ti_2O_7$ have been discussed in the literature up to now.

The crystal lattice of terbium titanate with the pyrochlore structure belongs to the space symmetry group Fd3m (face-centered cubic lattice). $Tb^{3+}$ ions are in the Wyckoff position 16d (1/2, 1/2, 1/2), $Ti^{4+}$ ions are in the position 16c (0, 0, 0), two sublattices O1 and O2 of $O^{2-}$ ions are in the positions 8b (3/8, 3/8, 3/8) and 48f ($x_0$, 1/8, 1/8), respectively. $Tb^{3+}$ ions form a network of tetrahedra connected by their vertices. Four $Tb^{3+}$ ions in a unit cell are crystallographically equivalent, but magnetically nonequivalent. The eight oxygen ions closest to the terbium ion form a strongly distorted cube. The local point symmetry group of the $Tb^{3+}$ ion is trigonal $D_{3d}$.

In the crystal field of a perfect $Tb_2Ti_2O_7$ crystal, the ground multiplet $^7F_6$ of the $Tb^{3+}$ ion is split into 4 doublets $E_g$ and 5 singlets $3A_{1g}+2A_{2g}$ (irreducible representations of the $D_{3d}$ group are shown). The ground and first excited states are non-Kramers doublets. The first excited energy level of the $Tb^{3+}$ ion lies 13 cm$^{-1}$ (0.39 THz) above the ground level [4, 5]. The remaining energy levels lie above 80 cm$^{-1}$. $Tb_2Ti_2O_7$ stands out among other isostructural rare earth compounds by the small energy gap between the ground and the first excited sublevel of the ground multiplet. In other rare earth pyrochlores, the first excited energy level is separated from the ground by an energy gap of more than 70 cm$^{-1}$. Observations of extremely large values of magnetoelastic effects in $Tb_2Ti_2O_7$ at low temperatures, in particular, giant forced magnetostriction [6] and coupled electron-phonon excitations [7, 8], indicate an anomalously strong electron-lattice interaction. At present, it is generally accepted in the literature that it is the combination of these factors and the competition of ferromagnetic and antiferromagnetic interactions between neighboring $Tb^{3+}$ ions that cause the absence of magnetic ordering in $Tb_2Ti_2O_7$ at low temperatures [9].

In the low-energy part of the inelastic neutron scattering spectra of $Tb_2Ti_2O_7$ single crystals and powders, an intense line is observed in the energy range of ~ 1 cm$^{-1}$ [10, 11]. In the terahertz absorption spectra, a broad line is observed in the frequency range corresponding to energies of 10−18 cm$^{-1}$ [12-14]. The specific profile of the inelastic neutron scattering spectrum in the region of low transfer energies (~ 1 cm$^{-1}$) and the fine structure of the terahertz absorption line indicate splitting of the ground and first excited non-Kramers doublets of the $Tb^{3+}$ion. The main objective of this work is to clarify the mechanism of formation of these splittings.

The observed specific features of low-temperature spectral, magnetic and thermodynamic properties of $Tb_2Ti_2O_7$ powders and single crystals (absence of magnetic ordering, dependence of heat capacity and magnetic susceptibility on the sample [15, 16], violations of selection rules and additional lines in optical spectra [17], Raman spectra and inelastic neutron scattering spectra) indicate the formation of microscopic inhomogeneities of the crystal lattice. Local static lattice deformations are induced during the synthesis of samples due to the formation of point defects (oxygen vacancies, violations of the stoichiometric composition [18, 19] during the substitution of $Ti^{4+}$ ions by $Tb^{3+}$ ions and $Tb^{3+}$ ions by $Ti^{4+}$ ions, characterized by the parameter $x$ in the formula of the real compound $Tb_{2+x}Ti_{2-x}O_{7-y}$).

It should be noted that the interpretation of the specific shape of the terahertz absorption line at the transition between the lower non-Kramers doublets of $Tb^{3+}$ ions proposed in [12–14], based on the consideration of coupled electron-phonon excitations, has no physical justification, since the hybridization of electron excitations within the states of the ground electron configuration with excitations of the cubic lattice (with phonons from the acoustic and odd optical branches of the vibrational spectrum) is possible (by parity) only at finite values of the phonon wave vectors, significantly exceeding the wave vectors of terahertz photons, which are close to zero.

**Crystal structure defects and the field of random deformations**

In the elastic continuum approximation, the structure of the crystal lattice at a distance $r$ from a point defect (displacements of ions from equilibrium positions $\boldsymbol{u}(\boldsymbol{r}) \sim 1/r^2$ [20]) is described by a non-uniform deformation tensor with components $e_{\alpha\beta}(\boldsymbol{r}) = e_{\beta\alpha}(\boldsymbol{r}) \sim 1/r^3$. In the case of a finite concentration of defects, the components of the strain tensor are random variables with the probability distribution density of strains $g(\boldsymbol{e})$. In the general case, the structure of the multidimensional distribution function $g(\boldsymbol{e})$ is determined by the symmetry properties of the real crystal. When considering crystals of cubic symmetry, in particular, pyrochlores, the six-dimensional space of the components of the strain tensor can be decomposed into subspaces corresponding to irreducible representations $A_{1g}$, $E_g$ and $F_{2g}$ of the cubic symmetry group $O_h$:

$$e_1 = e(A_g) = (e_{xx} + e_{yy} + e_{zz})/\sqrt{6},$$

$$e_2 = e(E_g, 1) = (2e_{zz} - e_{xx} - e_{yy})/\sqrt{12},$$

$$e_3 = e(E_g, 2) = (e_{xx} - e_{yy})/2,$$

$$e_4 = e(F_{2g}, 1) = (2e_{xy} - e_{xz} - e_{yz})/\sqrt{6},$$

$$e_5 = e(F_{2g}, 2) = (e_{xz} - e_{yz})/\sqrt{2},$$

$$e_6 = e(F_{2g}, 3) = (e_{xy} + e_{xz} + e_{yz})/\sqrt{3}. \qquad (1)$$

In an elastically isotropic continuum, the trace of the strain tensor induced by point defects is zero [20], and the distribution function takes the form [21, 22]:

$$g(\boldsymbol{e}) = \frac{2\gamma}{\pi^3}\left(\sum_{m=2}^{6} e_m^2 + \gamma^2\right)^{-3}. \qquad (2)$$

The width of the distribution is determined by the parameter $\gamma = \frac{\pi(1+\sigma)}{27(1-\sigma)} C_d |\Omega_0|$, where $C_d$ is the number of defects per unit volume, $\Omega_0$ is the "defect strength" and $\sigma$ is the Poisson ratio. The use of the distribution function (2) in the calculations of the inelastic neutron scattering spectrum profile in the region of low transfer energies in $Tb_2Ti_2O_7$ [23] and $Pr_2Zr_2O_7$ [24] crystals, the temperature dependence of the heat capacity of the $Pr_2Zr_2O_7$ crystal [24] at low temperatures made it possible to successfully reproduce the measurement data. The distribution function of

deformations in an elastically anisotropic continuum, as was shown in [25], can be approximated by a generalized 6-dimensional Lorentz function:

$$g(e) = \frac{15\xi}{8\pi^3 \gamma_A \gamma_E^2 \gamma_F^3} \left[ \frac{e(A_{1g})^2}{\gamma_A^2} + \sum_{\lambda=1}^{2} \frac{e(E_g,\lambda)^2}{\gamma_E^2} + \sum_{\lambda=1}^{3} \frac{e(F_{2g},\lambda)^2}{\gamma_F^2} + \xi^2 \right]^{-7/2}. \quad (3)$$

It should be noted that the width of the distributions of deformations of different symmetries depends on temperature, since the ratios between the measured elastic constants of the $Tb_2Ti_2O_7$ crystal change significantly in the low-temperature region ($T < 80$ K) [26,27]. Distribution function (3) was used in the calculations of the spectral and thermodynamic characteristics of the $Pr_2Zr_2O_7$ crystal in [28]. In the case of a high concentration of defects, the distribution function of deformations in an elastically isotropic continuum is transformed and takes the form of a Gaussian distribution [21]. In particular, the Gaussian distribution function of random strains has been used in studies of the cooperative Jahn-Teller effects in solid solutions of rare-earth vanadates in [29]. In the present paper, the distribution function of deformations induced by point defects in the $Tb_2Ti_2O_7$ crystal is approximated by a generalized Gaussian distribution:

$$g(e) = \frac{1}{\pi^3 \xi^6 \gamma_A \gamma_E^2 \gamma_F^3} \exp\left\{ -\left[ \frac{e(A_{1g})^2}{(\gamma_A \xi)^2} + \sum_{\lambda=1}^{2} \frac{e(E_g,\lambda)^2}{(\gamma_E \xi)^2} + \sum_{\lambda=1}^{3} \frac{e(F_{2g},\lambda)^2}{(\gamma_F \xi)^2} \right] \right\}. \quad (4)$$

The ratio of the widths of the distributions of deformations of different symmetries is taken to be the same as in the case of the Lorentz distribution (parameters $\gamma_A = 1.2$, $\gamma_E = 31.8$, $\gamma_F = 24.8$ were calculated using the elastic constants of $Tb_2Ti_2O_7$ measured at a temperature of $T = 6$ K [30]). The parameter $\xi$, depending on the type and concentration of defects, was a fitting variable.

**Crystal field and electron-deformation interaction**

Calculations of energies of the Stark sublevels of $Tb^{3+}$ ions in $Tb_2Ti_2O_7$ were performed using the numerical diagonalization of the Hamiltonian

$$H = H_{FI} + H_{CF}, \quad (5)$$

defined in the full space of 3003 states of the electronic $4f^8$ configuration. Here $H_{FI}$ is the standard Hamiltonian of a free ion [31], $H_{CF}$ is the operator of interaction of 4f-electrons with the static crystal field in the regular crystal lattice.

The Hamiltonian $H_{CF}$ of four magnetically nonequivalent $Tb^{3+}$ ions in local Cartesian coordinate systems with $Z$ axes along the local $C_3$ axes and $X$ axes in the planes containing the $Z$ axis and one of the crystallographic $C_4$ axes is determined by six non-zero real parameters $B_p^k$:

$$H_{CF} = B_2^0 O_2^0 + B_4^0 O_4^0 + B_4^3 O_4^3 + B_6^0 O_6^0 + B_6^3 O_6^3 + B_6^6 O_6^6. \quad (6)$$

Here, $O_p^k$ are linear combinations of spherical tensor operators, similar to Stevens operators [32] in the space of eigenfunctions of the angular momentum operator. Crystal-field parameters $B_2^0$=213, $B_4^0$=368, $B_4^3$=–2574, $B_6^0$=49.7, $B_6^3$=1218, $B_6^6$=1002 (cm$^{-1}$) were calculated within the framework of the exchange charge model [33] (the ion charges used in the calculation equal $q_{Tb} = +2.82$, $q_{Ti} = +3.58$, $q_{O1} = -1.64$, $q_{O2} = -1.86$ (in units of elementary charge), parameters of charges on the bonds Tb–O1 and Tb–O2 equal $G_\sigma = G_s = G_\pi = 6.9$ and $G_\sigma = G_s = G_\pi = 10.5$, correspondingly.

Table 1. Calculated and measured energies of the Stark sublevels of the ground multiplet $^7F_6$ of Tb$^{3+}$ ions in Tb$_2$Ti$_2$O$_7$ (cm$^{-1}$), the corresponding irreducible representations of the group $D_{3d}$ are indicated.

|  | $E_g$ | $E_g$ | $A_{2g}$ | $A_{1g}$ | $E_g$ | $A_{2g}$ | $A_{1g}$ | $E_g$ | $A_{1g}$ |
|---|---|---|---|---|---|---|---|---|---|
| Theory | 0 | 11.6 | 82 | 137 | 311 | 385 | 389 | 484 | 566 |
| Experiment [4] | 0 | 13 | 83 | 135 | – | – | – | – | – |
| Experiment [5] | 0 | 12.1 | 82 | 135 | 339 | – | 395 | 492 | – |

The energies of the sublevels of the ground multiplet (see Table 1) and the *g*-factors of the ground ($g_\parallel$ = 10.5) and first excited ($g_\parallel$ = 13.6) doublets obtained using the calculated crystal-field parameters are in good agreement with the data of measurements of the IR absorption spectra [4] and inelastic neutron scattering [5] and magnetization of Tb$_2$Ti$_2$O$_7$ in external magnetic fields [34].

A point defect locally deforms the crystal lattice, the equilibrium positions of ions $R(L,s)$ (*L* is the unit cell number and *s* is the ion number in the unit cell) shift by vectors $u(R,s)$. As in the case of a uniformly deformed crystal, when considering a field of random deformations, we can introduce a deformation tensor with components $e_{\alpha\beta}$ and sublattice displacements $w(s)$: $u_\alpha(L,s) = \sum_\beta e_{\alpha\beta} X_\beta(L,s) + w_\alpha(s)$. The crystal field parameters undergo increments $\Delta B_p^k = B_p^k(R+u) - B_p^k(R)$. Note, we neglect here odd components of the crystal field induced by point defects which, in the case of local $D_{3d}$ symmetry, can affect physical properties of rare-earth ions in the second order of the perturbation theory. However, odd local static lattice deformations similarly to odd vibrational lattice modes lift the ban on electric-dipole radiative transitions.

In the linear approximation in the components of the deformation tensor and the displacement vectors of the sublattices, the Hamiltonian describing the interaction of the Tb$^{3+}$ ion with lattice deformations has the form [33]:

$$H_{\text{el-def}} = \sum_{pk}[\sum_{\alpha\beta} B'^k_{p,\alpha\beta} e_{\alpha\beta} + \sum_{\alpha,s} B''^k_{p,\alpha}(s) w_\alpha(s)] O^k_p, \quad (7)$$

where

$$B'^k_{p,\alpha\beta} = \frac{1}{2}\sum_{L,s}\left[X_\alpha(L,s)\frac{\partial}{\partial X_\beta(L,s)} + X_\beta(L,s)\frac{\partial}{\partial X_\alpha(L,s)}\right] B^k_p, \quad (8)$$

$$B''^k_{p,\alpha}(s) = \sum_L \frac{\partial B^k_p}{\partial X_\alpha(L,s)}. \quad (9)$$

Table 2. Parameters of the Hamiltonian of electron-deformation interaction (cm$^{-1}$).

| $p, k$ | $\tilde{B}(A^2_{1g})$ | $pk$ | $\tilde{B}^k_p(E^1_g,1)$ | $\tilde{B}^k_p(E^2_g,1)$ | $p, k$ | $\tilde{B}^k_p(E^1_g,2)$ | $\tilde{B}^k_p(E^2_g,2)$ |
|---|---|---|---|---|---|---|---|
| 2, 0 | −8921 | 2, 1 | 10140 | −8673 | 2, −1 | 10140 | 8673 |
| 4, 0 | −1882 | 4, 1 | 2311 | −5686 | 4, −1 | 2311 | 5686 |
| 4, 3 | −14621 | 6, 1 | 2470 | 340 | 6, −1 | 2470 | −340 |
| 6, 0 | −483 | 4, 4 | −2881 | −7954 | 4, −4 | −2881 | 7954 |
| 6, 3 | 4242 | 6, 4 | −229 | 2077 | 6, −4 | −229 | −2077 |
| 6, 6 | 3546 | 2, 2 | 7501 | −10960 | 2, −2 | −7501 | −10960 |
| | | 4, 2 | −2188 | 2760 | 4, −2 | 2188 | 2760 |
| | | 6, 2 | 709 | −982 | 6, −2 | −709 | −982 |
| | | 6, 5 | −615 | 3700 | 6, −5 | 615 | 3700 |

The influence of sublattice displacements can be taken into account by renormalizing the parameters $B^{k'}_{p,\alpha\beta}$, and the Hamiltonian $H_{\text{el-def}}$ takes the form

$$H_{\text{el-def}} = \sum_{pk,\alpha\beta}\tilde{B}'^k_{p,\alpha\beta} e_{\alpha\beta} O^k_p = \sum_{pk,\Gamma\lambda}\tilde{B}'^k_p(\Gamma\lambda) e(\Gamma\lambda) O^k_p. \quad (10)$$

Renormalized parameters $\tilde{B}'^k_p(\Gamma\lambda)$ of coupling with symmetrized components of the strain tensor $e(\Gamma\lambda)$, transforming along a row $\lambda$ of the irreducible representation $\Gamma$, were obtained by us in [34] within the framework of the exchange charge model with subsequent correction for better agreement with experimental data on the field dependence of forced magnetostriction and the temperature dependences of elastic constants. In the local coordinate system at the position of Tb$^{3+}$ ions with the symmetry axis C$_3$∥[111], the Hamiltonian of the electron-deformation interaction takes the form $H_{\text{el-def}} = \sum_{\Gamma\lambda} V(\Gamma\lambda)\varepsilon(\Gamma\lambda)$, where $\varepsilon(\Gamma\lambda)$ are linear combinations of the deformation tensor components transforming in accordance with irreducible representations of the $D_{3d}$ group, $\varepsilon(A_{1g}^1) = e(A_{1g})$, $\varepsilon(A_{1g}^2) = e(F_{2g},3)$, $\varepsilon(E_g^1,1) = e(E_g,1)$, $\varepsilon(E_g^1,2) = e(E_g,2)$, $\varepsilon(E_g^2,1) = e(F_{2g},1)$, $\varepsilon(E_g^2,2) = e(F_{2g},2)$, and operators $V(\Gamma\lambda) = \sum_{p,k} \tilde{B}_p^k(\Gamma\lambda) O_p^k$. The parameters of the electron-deformation interaction used in the present work are given in Table 2.

**Absorption spectrum profile**

The absorption spectra of linearly polarized synchrotron radiation in the frequency range of 0.2–1 THz at temperatures of 6–300 K were measured on single-crystal samples of $Tb_{2+x}Ti_{2-x}O_{7-y}$ grown by the zone melting method in [14]. The radiation was directed along one of the trigonal ([111]) or rhombic ([1-10]) symmetry axes of the crystal with an alternating magnetic field **h** ∥ [11-2], **h** ∥ [1-10] or **h** ∥ [111], respectively, in crystallographic system of coordinates.

A comparison of the results of measurements of the low-temperature heat capacity of the samples used in spectroscopic studies [12] with the temperature dependences of the heat capacity of terbium titanate samples with different relative contents of $Tb^{3+}$ and $Ti^{4+}$ ions presented earlier in the literature [35] enabled the authors of [12–14] to estimate the value of the parameter $x$, which characterizes the deviation from the stoichiometric composition, for samples with significantly different absorption spectra caused by electromagnetic radiation-induced transitions of $Tb^{3+}$ from the ground state (doublet $E_g^1$) to the nearest excited state (doublet $E_g^2$), the transition frequency in the crystal field being ~0.43 THz. In particular, in samples whose absorption spectrum at low temperatures contains a broad line of complex shape, corresponding to the transition $E_g^1 \Rightarrow E_g^2$, and an additional weak line of unknown nature at a frequency of 0.67 THz, parameter $x \approx -0.0025$. Samples with $x>0$ and $x<0$ contain defects of various types. In crystals with $x>0$, heterovalent substitution of $Ti^{4+}$ ions by $Tb^{3+}$ ions is accompanied by the appearance of vacancies in the oxygen sublattices; at $x<0$, filling of the 8a positions by oxygen ions is possible.

Calculations of the absorption spectra of four magnetically nonequivalent $Tb^{3+}$ ions at the vertices of the tetrahedron were performed based on the consideration of the single-ion Hamiltonian in the corresponding local coordinate systems

$$H_t = H + H_{\text{el-def}} + H_r, \qquad (11)$$

where $H_{\text{el-def}}$ is the above-introduced operator of interaction of 4f-electrons with the field of random deformations, and $H_r$ is the operator of interaction with radiation. Since zero-phonon optical electric dipole transitions between states of the electronic $4f^8$ shell in the crystal field of $D_{3d}$ symmetry are forbidden by parity, in the operator $H_r = -\mathbf{M}\cdot\mathbf{h}$ ($\mathbf{M} = -\mu_B(k\mathbf{L}+2\mathbf{S})$ is the magnetic moment operator of a $Tb^{3+}$ ion, $\mathbf{L}$ and $\mathbf{S}$ are the total orbital and spin moments, respectively, $\mu_B$ is the Bohr magneton, $k$=0.95 is the orbital reduction factor) interactions with the magnetic field of radiarion are accounted for only.

In the general case, lattice deformations of $E_g$ symmetry split each of the $E_g^1$ and $E_g^2$ doublets into two sublevels with energies $E_1(\mathbf{e})$ and $E_2(\mathbf{e})$, $E_3(\mathbf{e})$ and $E_4(\mathbf{e})$, respectively, and wave functions $|i(\mathbf{e})>$ ($i$= 1-4); fully symmetric deformations shift the upper doublet relative to the lower one. Thus, at low temperatures in the space of four states of the two lower doublets there are six absorption channels, to which six spectral lines correspond: P0 ($E_1 \Rightarrow E_2$), P1 ($E_2 \Rightarrow E_3$), P2 ($E_2 \Rightarrow E_4$), P3 ($E_1 \Rightarrow E_3$), P4 ($E_1 \Rightarrow E_4$), P5 ($E_3 \Rightarrow E_4$).

Considering the operator $H_r$ as a time-dependent perturbation, we obtain the distribution of the absorption intensity of radiation of frequency $\omega$, caused by transitions between sublevels of terbium ions numbered with index $s$, with energies $E_{i,s}(\mathbf{e})$ and $E_{k,s}(\mathbf{e})$ for the fixed components of the deformation tensor $\mathbf{e}$, in the form

$$I(\omega,\mathbf{e}) = C\omega \sum_{s=1}^{4}\sum_{i=1}^{2}\sum_{k=3}^{4}[p_i(\mathbf{e})-p_k(\mathbf{e})]|\langle i|\mathbf{M}_s\cdot\mathbf{h}_{0,s}|k\rangle|^2 I_{0,ik}(\omega,\mathbf{e}), \qquad (12)$$

where $C$ is the numerical factor, $p_i$ is the population of the $i$-th state, $\mathbf{M}_s$ is the magnetic moment operator and $\mathbf{h}_{0,s}$ is the unit vector directed along the polarized radiation magnetic field in the local system of coorfinstes for the ion $s$, respectively,

$$I_{0,ik}(\omega,\mathbf{e}) = \left(2\pi\sigma^2\right)^{-1/2}\exp\left\{-\left[\omega-(E_k(\mathbf{e})-E_i(\mathbf{e}))/\hbar\right]^2/2\sigma^2\right\} \qquad (13)$$

is the Gaussian form-function of the spectral line corresponding to a separate transition with the dispersion $\sigma$ =7.5·10$^{-3}$ THz.

The absorption spectrum profile is obtained by averaging the distribution (12) with the deformation distribution function (4):

$$I(\omega) = \int I(\omega,\mathbf{e})g(\mathbf{e})de(A_{1g})\prod_{\lambda=1,2}de(E_g,\lambda)\prod_{\lambda=1,2,3}de(F_{2g},\lambda). \qquad (14)$$

Results of numerical calculations of integrals (14) in a hyperspherical coordinate system in a six-dimensional space of strain tensor components with a frequency step $\Delta\omega = 3.75\cdot10^{-3}$ Hz ($T$ = 6 K, $\mathbf{h}\|[11\text{-}2]$) are compared with the measured absorption spectrum ([14], Fig. 3S) in Fig. 1. The calculation of the absorption spectrum profile is performed under the assumption that the intensity of the terahertz radiation source does not depend on the frequency. The value of the variable parameter $\xi$ = 6.5·10$^{-4}$, as well as the presented above standard deviation $\sigma$ in the form-function of optical transitions (13), was determined by comparing the calculated spectrum with the measured one in [14]. Figure 1 also shows the lines P1, P2, P3 and P4, the superposition of which determines the shape of the envelope of magnetic dipole transitions between the sublevels of the ground and first excited doublets of $Tb^{3+}$ ions in a crystal field, split by random deformations. The optical transitions that form the lines P2 and P3, P1 and P4 are induced predominantly by projections of the field $\boldsymbol{h}$ onto the local planes $XY$ and the $Z$ axis, respectively. The ratio of the calculated sums of the intensities of the specified lines $I(P2+P3)/I(P1+P4)=1.4$ agrees qualitatively with the ratio of the squares of the modules of the matrix elements of the projections of the magnetic moment on the $X$ and $Z$ axes on the wave functions of the doublets $E_g^1$ and $E_g^2$ in the crystal field of the regular lattice $|M_X(1,3)/M_Z(1,4)|^2=1.7$ (the matrix elements of the operators $M_X$ and $M_Y$ are equal in module), however, the calculations performed do not reproduce the dependence of the absorption spectrum profile on the orientation of the field $\boldsymbol{h}$ in the crystallographic coordinate system observed in the measured spectra.

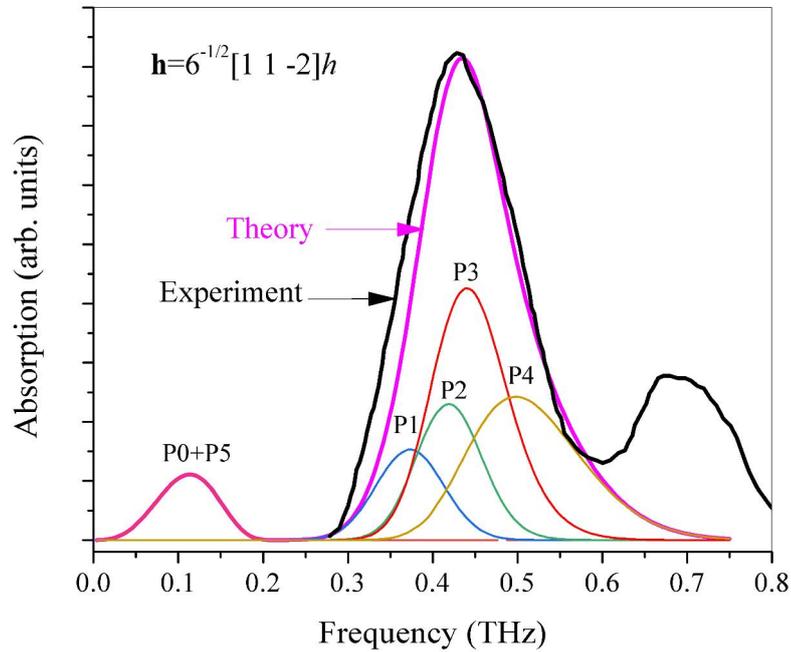

Fig. 1. Calculated absorption spectrum profile of linearly polarized radiation ($\boldsymbol{h}=2^{-1/2}[1-10]h$) of the $Tb_2Ti_2O_7$ single-crystal in comparison with measurement data [14] at a temperature of 6 K.

An additional line in the frequency range $0.7\pm0.05$ ТГц, observed in the measured spectrum, corresponds to the absorption of radiation by $Tb^{3+}$ ions located either in $Ti^{4+}$ positions or closest to a point defect that strongly changes the crystal field in neighboring lattice sites.

The P0+P5 line in Fig. 1 with the P0 line contribution dominating at low temperatures represents the envelope of transitions between the sublevels of $E_g^1$ and $E_g^2$ doublets split by random deformations. It should be noted that the splitting of the doublet is reproduced by calculation only when using a multidimensional (at least two-dimensional when considering deformations of $E_g$ symmetry [21]) distribution function with the most probable non-zero value of the vector modulus with components equal to the degenerate deformations, $|\mathbf{e}(\Gamma)| = [\sum_\lambda e(E_g\lambda)^2]^{1/2}$, $\Gamma = E_g, F_{2g}$.

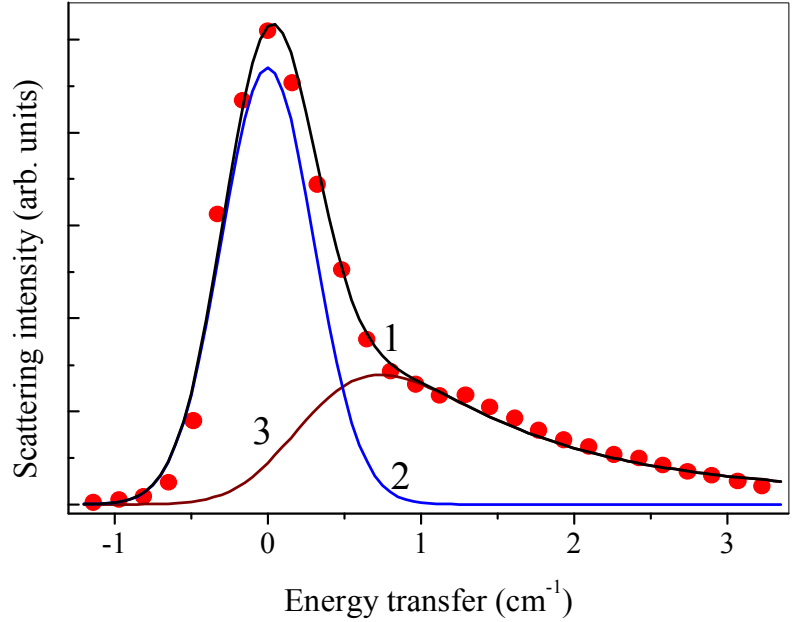

Fig. 2. Measured (symbols) [10] and calculated (line 1) neutron scattering spectra of the $Tb_2Ti_2O_7$ crystal at $T$=0.07 K. Lines 2 and 3 correspond to the contributions of elastic and inelastic scattering.

The splitting of the ground doublet $\Delta$ of $Tb^{3+}$ ions in $Tb_2Ti_2O_7$ crystals of an order of 1-2 cm$^{-1}$ was recorded in high-resolution inelastic neutron scattering spectra [10, 36] at low temperatures. The presence of a broad tail of the intense line in the region of quasi-elastic neutron scattering with transfer energies close to zero (see Fig. 2) indicates a Lorentzian distribution of random deformations in the studied samples, apparently with a small deviation from the stoichiometric composition. In this paper, the intensity distribution in the inelastic neutron scattering spectrum averaged over scattering vectors

$$I_{\text{inelast}}(E) \sim \int \exp\left[-(E - \Delta(\mathbf{e}))^2 / 2\delta^2\right] g(\mathbf{e}) d\mathbf{e} \qquad (15)$$

was calculated using the strain distribution function (2) in the form of a generalized Lorentz function with the width $\gamma = 5.5 \cdot 10^{-4}$, the dependence of the splitting of the doublet $\Delta(e)$ on the components of the strain tensor was obtained using the considered above Hamiltonian of the electron-deformation interaction. The observed neutron scattering spectrum profile is well reproduced by the sum of the intensities of the inelastic scattering and elastic scattering spectra, in which the intensity distribution was approximated by a Gaussian form-function $I_{\text{elast}}(E) \sim \exp(-E^2/2\delta^2)$ with the standard deviation $\delta = 0.3$ cm$^{-1}$ (see Fig. 2).

It should be noted that practically similar results of the calculation of the inelastic neutron scattering spectrum were obtained by taking into account the elastic anisotropy of the terbium titanate lattice using the random deformation distribution function (3) with a width of $\xi = 2.6 \cdot 10^{-5}$.

**Conclusions**

In this work, we calculated the envelopes of low-temperature absorption spectra of linearly polarized terahertz radiation and inelastic neutron scattering in the absence of external magnetic fields in Tb$_{2+x}$Ti$_{2-x}$O$_{7-y}$ crystals, taking into account the fields of random deformations induced by point defects of the crystal lattice. Modeling the spectra within the single-ion approximation allowed us to satisfactorily describe the presented in the literature measured absorption profiles of radiation with the wave vector *q*||[111] and magnetic field *h*||[11-2] in the crystal with a composition different from stoichiometric one ($x = -0.0025$) and a fragment of the neutron scattering spectrum in the region of low transfer energies (1 cm$^{-1}$) in a crystal with the parameter $x \sim 0$. The spectral lines considered correspond to quantum transitions between the sublevels of the two lower non-Kramers doublets of Tb$^{3+}$ ions in the crystal field of $D_{3d}$ symmetry, split by random deformations, in the absorption spectra and between the sublevels of the ground doublet in the neutron scattering spectrum. The width of the deformation distribution function ($6.5 \cdot 10^{-4}$ and $2.6 \cdot 10^{-5}$ in crystals of different composition), found from a comparison of the calculated and measured spectra, agrees with the previously obtained estimates of the width of the deformation distribution functions in dielectric crystals activated by rare-earth ions [22, 25].

In order to reproduce the dependence of the relative intensities of the four components of the broad absorption line on the polarization of terahertz radiation, it is necessary to take into account the possibility of lifting the ban on electric dipole transitions due to displacements of Tb$^{3+}$ ions from the nodes of the regular lattice near defects, accompanied by the appearance of an odd component of the crystal field, and to go beyond the single-particle approximation by including anisotropic exchange [38], magnetic dipole-dipole and electric multipole (in particular, quadrupole [39]) interactions between $Tb^{3+}$ ions in the Hamiltonian of the system under consideration.

This work was funded by the subsidy allocated to Kazan Federal University for thestate assignment in the sphere of scientific activities (Project No. FZSM-2024-0010).